\documentstyle[sprocl]{article}
\input{epsf}
\def\Journal#1#2#3#4{{#1}{\bf #2}, #3 (#4)}

\def\PLB{{\em Phys.~Lett.}~B}
\def\PRL{\em Phys.~Rev.~Lett.~}
\def\PRD{{\em Phys.~Rev.}~D}

\def\be{\begin{equation}}
\def\ee{\end{equation}}
\def\bea{\begin{eqnarray}}
\def\eea{\end{eqnarray}}
\def\psl{\hbox{/\kern-.5600em$p$}}
\def\darr{\raise1.5ex\hbox{$\leftrightarrow$}\mkern-16.5mu \partial_\mu} 

\def\ze{\phantom{-}0}

\begin{document}
\rightline{McGill/96-17}
\rightline{hep-ph/9606232}
\rightline{May, 1996}
\vskip .4in
\title{$R_b$ PROBLEM: LOOP CONTRIBUTIONS AND SUPERSYMMETRY}
\author{J.M.~CLINE}
\address{McGill University, 3600 University St., Montr\'eal,
Qu\'ebec H3A 2T8, Canada}

\maketitle\abstracts{I summarize a part of the work done with Bamert,
Burgess, London and Nardi,\cite{BBCLN}\ in which we find simplified
expressions for the possible one-loop contributions to the $R_b$
parameter from new virtual particles, including supersymmetric ones.
Our expressions make it easier to identify which models and choices of
parameters are best able to solve the $R_b$ problem.}

The evidence for a discrepancy between the standard model prediction
for $R_b$, the ratio of partial decay widths for $Z\to b\bar b$ to
$Z\to$ hadrons, has been described by P.~Bamert.\cite{PB}\  In this
talk I will recapitulate our investigation$\,$\cite{BBCLN}\ of the
possibility of resolving the $R_b$ problem using new particles in the
loops of the diagrams of Figure 1, which contribute to the $Z\to b\bar
b$ width.  For example, possible combinations of particles in the loop
could be an extra Higgs boson and the top quark, a squark and a
chargino, or the usual Higgs with a 4th generation quark.  Many
different extensions of the standard model have been suggested for
solving the $R_b$ problem.\cite{models}\  However, it would be nice if
it was not necessary to compute the corrections to $R_b$ anew for every
such possibility.  Our goal in this study is to identify what are the
new physics ingredients needed to increase the value of $R_b$ through
one-loop effects, in as model-independent a way as possible.

\section{Methodology}

\begin{figure}[t] 
\epsfysize=1in\epsfbox{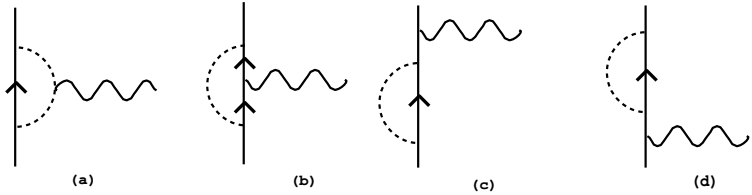} 
\caption{Loop corrections to the $Z\to b\bar b$ decay width.
\label{fig:loops}}
\end{figure}

\begin{figure}[b] 
\epsfysize=0.8in\epsfbox{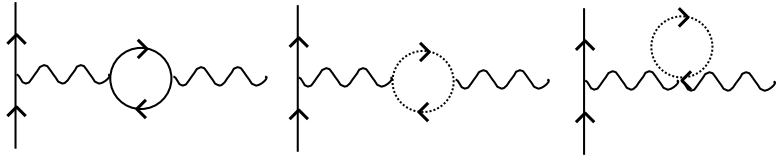} 
\caption{Vacuum polarization corrections to the $Z\to b\bar b$ decay width,
which we ignore.
\label{fig:vp}}
\end{figure}

To achieve our goal, it is very useful to make three simplifications.

(1) Ignore corrections to the $Z b_R\bar b_R$ vertex.  The left- and
right-handed $b$ quarks couple to the $Z$ boson through the interaction
\begin{equation}
	\frac12 \bar b \gamma^\mu\left( g^b_L (1-\gamma_5) + 
	g^b_R (1+\gamma_5) \right) b\, Z_\mu.
\label{bcouplings}
\end{equation}
As already discussed,\cite{PB}\ the tree-level value of $g^b_R$ is so
small that we would need to increase it by 100\% if we wanted to solve
the $R_b$ problem just by changing $g^b_R$.  Such a large shift would
be quite difficult to achieve from loop corrections for any
reasonable-size coupling constants.  On the other hand, to solve the
$R_b$ problem by increasing $g^b_L$ is possible at one loop, as is
demonstrated by the fact that the absence of the top quark loop
correction to $R_b$ in the standard model would be enough to remove the
discrepancy.

(2) Set the external momentum $q$ of the $Z$ boson to $q^2=0$ rather than
its actual value of $q^2=m_Z^2$.  This makes the formulas for the loop
diagrams much more tractable, and typically introduces only a small error,
as I will show below.

(3) Ignore vacuum polarization corrections, as in figure 2.  At
one-loop accuracy we are justified in doing so, because they largely
cancel in the ratio which defines $R_b$.  Let us denote the relative
corrections to the hadronic widths due to the vacuum polarization
diagrams of figure 2 by $\epsilon_{vp}$, and the absolute corrections
from figure 1
to the partial widths into $b\bar b$ and hadrons, respectively, by
$\epsilon_b$ and $\epsilon_h$:
\begin{equation}
	R_b \equiv {\Gamma_b\over\Gamma_{\rm hadrons}} =
	{\Gamma_b^0 (1 + \epsilon_{vp}) + \epsilon_b \over
	\Gamma_h^0 (1 + \epsilon_{vp}) + \epsilon_h }
	= {\Gamma_b^0  + \epsilon_b + {\cal O}(\epsilon_b\epsilon_{vp})
	\over
	\Gamma_h^0  + \epsilon_h + {\cal O}(\epsilon_b\epsilon_{vp})},
\end{equation}
where $\Gamma^0_i$ is the tree-level value.  Because $\epsilon_{vp}$
is a universal correction for all flavors of quarks, it cancels in the ratio,
to one loop accuracy. 

This last simplification is helpful not only because it reduces the
number of diagrams that must be calculated, but also because the ones
that remain (figure 1) have nice properties.  Thanks to the Ward
identity, the diagrams 1a and 1b are related to those of 1c and 1d in
such a way that the sum of the four is finite and gauge invariant.
Let us recall how this works.  If $p'$ and $p$ are the initial and final
momenta of the $b$ quark, respectively, then the Ward identity is
\bea
	(p-p')^\mu \Gamma_\mu &=& g\left(S_F^{-1}(p) - S_F^{-1}(p')\right);
	\nonumber\\
	(g + \delta g) (\psl - \psl') &=& g(1+\delta Z) (\psl - \psl').
\eea
Here $\Gamma_\mu$ is the 1PI contribution to the vertex correction,
which can be written as $(g + \delta g)\gamma_\mu$, in terms of the
tree-level coupling $g$ and its one-loop shift $\delta g$, and $\delta
Z$ is the wave function renormalization.  Thus diagrams 1a and 1b are
associated with  $\delta g$ and similarly 1c and 1d with $\delta Z$.
Although the Ward identity is derived for an unbroken $U(1)$ symmetry,
the divergent parts of the diagrams are the same for 
the $Z$ boson as for the photon, so the sum of the diagrams is finite.
Furthermore, we shall see that in our approximation of $q^2=0$ (point 2
above), the parts of the diagrams proportional to $\sin^2\!\theta_W$ combine
to give zero.  Again, these are the parts proportional to the photon coupling,
and the neglect of the $Z$ mass makes the corresponding U(1) symmetry appear
to be unbroken.

\section{Results}

Before presenting the results for the loop diagrams, here are our
conventions for the couplings of the new fermion ($f$) and scalar
($\phi$) in the loop to the $Z$ boson and to each other:
\bea
{\cal L}_{\rm gauge} &=&
\frac{e}{\sin\! 2\theta_W} \left[ \bar f \gamma^\mu\left( g^f_L (1-\gamma_5) 
+ g^f_R
(1+\gamma_5) \right) f + 2i g^\phi \phi^\dagger\darr\phi\right]\, Z_\mu;
\nonumber\\
{\cal L}_{\rm Yukawa} &=& {y\over 2} \bar b \phi (1+\gamma_5) f + {\rm h.c.}
\label{couplings}
\eea
Notice that we have not bothered to write a Yukawa coupling to $b_R$,
the right-handed $b$ quark, since we are only interested in generating
corrections to $g^b_L$.  Moreover gauge invariance demands the relation
$g^b_L = g\phi + g^f_L$.  The gauge couplings are related to the third
component of weak isospin and the electric charge in the usual way,
\be
	g = (T_3 -  Q\sin^2\!\theta_W).
\ee
It is these $\sin^2\!\theta_W$'s appearing in the gauge couplings that
cancel out of all the results in the limit of vanishing $m_Z$.

The one-loop shift in $g^b_L$ due to the diagrams of figure 1
now has a very simple form,
\be
	\delta g^b_L = {y^2 n_c\over 16 \pi^2} \left( g^f_L - g^f_R \right)
	{\cal F}(m^2_f/m^2_\phi); \qquad
	{\cal F}(r) = {r\over r-1} - {r\ln r\over (r-1)^2},
\label{result}
\ee
where $n_c$ is a color factor ({\it e.g.,} $n_c=2$ if the particles in
the loop are color triplets) and ${\cal F}$ is a positive function of
the ratio of the fermion to the scalar mass, with $0 \le {\cal F}\le
1$, as shown in figure 3.  Although the gauge coupling of the scalar,
$g^\phi$, appears in the calculation of $\delta g^b_L$, it can be
eliminated in favor of the chiral couplings of the fermion, due to
gauge invariance.  As advertised, the $\sin^2\!\theta_W$-dependence has
disappeared from the result, and $( g^f_L - g^f_R)$ depends only on the
isospins of the two chiral fermions in the loop.

\begin{figure} 
\epsfysize=3.0in\epsfbox[-300 0 612 792]{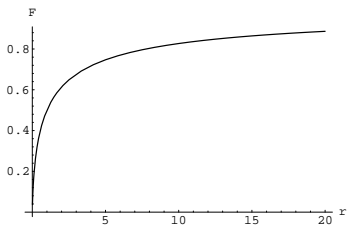} 
\vspace{-0.9in}
\caption{The function ${\cal F}$ of the fermion to scalar mass ratio from
eq.~(\ref{result}).
\label{fig:calF}}
\end{figure}

We can immediately discern from eq.~(\ref{result}) what is needed to
increase $R_b$: the combination $( g^f_L - g^f_R)$ must have the same
sign (negative) as $g^b_L$ itself.  Thus two-Higgs doublet models will
not work, to the extent that the top quark Yukawa coupling dominates in
the loop over that of the bottom, since $( g^t_L - g^t_R)$ has the
opposite sign to $g^b_L$.  This computation also shows why the large
top quark mass exacerbates the $R_b$ problem in the standard model
itself.

However, one might be concerned about our neglect of the mass of the $Z$
boson.  It is straightforward to make an expansion in powers of $m^2_Z$
over the other masses to see how big the error is.  For simplicity
consider the special case where the fermion and the scalar in the loop
are degenerate, $m_f = m_\phi$.  Then to first order in $m^2_Z/m^2_f$,
eq.~(\ref{result}) becomes
\be
	\delta g^b_L = {y^2 n_c\over 32 \pi^2} \left( g^f_L - g^f_R
	+{m^2_Z\over 12 m^2_f} (g^b_L+g^f_L+g^f_R)\right).	
\ee
Thus even if the new particles in the loop were as light as the $Z$
boson, the correction due to the nonvanishing $Z$ mass would still only
be $\sim$10\% of the leading order result.

\section{Nondiagonal $Z$-couplings and supersymmetry}

So far we assumed that the new particles in the loop consisted of a
single fermion and scalar, but it is possible that there are several of
each, and that their couplings to the $Z$ boson, although diagonal in the
flavor basis, are no longer diagonal in the basis of mass eigenstates.
For example two flavors of fermions $a$ and $b$ would have the chiral
rotation matrices $U_{L,R}$ defined by
\be
	\left( \begin{array}{c} f^a_{L,R} \\  f^a_{L,R} \end{array}\right)
	= \left( \begin{array}{rr} \phantom{-}
	\cos\theta_{L,R} & \sin\theta_{L,R}\\
	-\sin\theta_{L,R} & \cos\theta_{L,R} \end{array}\right)
	\left( \begin{array}{c} f^1_{L,R} \\  f^2_{L,R} \end{array}\right)
\ee
and similarly for the bosons.  This is just what happens in
supersymmetric models, where the fermions are the higgsino $\tilde
h^-_2$ which couples to the bottom quark and top squark with the top
quark Yukawa coupling $y_t$, and the Wino, $\widetilde W^-$. Their mass 
matrix, in the basis ($\tilde h^-_2$, $\widetilde W^-$), is
\be	m^2_{\rm chargino} =
   \left( \begin{array}{cc}\mu & g\,v_2 \\ g\,v_1 & M_2 \end{array}\right)
\label{chargino}
\ee
which is generally nonsymmetric, so that it is diagonalized by a similarity
transformation with $U_L\neq U_R$.  Likewise there are two bosons, the
right- and left-handed top squarks $\tilde t_{R,L}$ with mass matrix
\be m^2_{\rm squark} =
\left( \begin{array}{cc} m^2_U + m^2_t + {\cal O}(m^2_W) & 
m_t(A_t\sin\beta + \mu\cos\beta) \\ m_t(A_t\sin\beta + \mu\cos\beta)
 &  m^2_Q + m^2_t + {\cal O}(m^2_W)  \end{array}\right).
\ee
In the flavor basis, the gauge coupling constants of the previous section
become matrices,
\be
	g^f_L = g^f_R = \left( \begin{array}{cc} -\frac12 & \ze \\
	\ze & -1  \end{array}\right),\qquad g^\phi = 
  	\left( \begin{array}{cc} 0 & 0 \\
	0 & \frac12  \end{array}\right),
\label{charges}
\ee
where we ignored the $\sin^2\!\theta_W$ parts as justified above.

The generalized expression for $\delta g^b_L$ corresponding to
eq.~(\ref{result}) is much more complicated when we have multiple
particles with mixing in the loop.\footnote{For detailed formulas in the
case of the minimal supersymmetric standard model, see ref.~4.}
It can be written in terms of some elementary double integrals,
\bea
&&\delta g^b_L = {y^2 n_c\over 32\pi^2} \int_0^1\!\! du \int_{-u}^u\!\!\! 
dv
\sum_{ijkl} U^\phi_{1i} {U^{\phi}_{1j}}^* U^R_{1k} {U^R_{1l}}^* \Bigl\{
g^b_L \delta_{ij} \delta_{kl} \ln(u\, m^2_{\phi_i} + (1-u)m^2_{f_l})
\nonumber\\
&&-\, (U_R^\dagger g^f_R U_R)_{kl}\delta_{ij}\left( 1+ \ln\left( 
(1-u)m^2_{\phi_i} +\frac{u}{2}(m^2_{f_k}+m^2_{f_l}) - 
\frac{v}{2}(m^2_{f_k}-m^2_{f_l})\right)\right)\nonumber\\
&&+\, \delta_{ij} m_{f_k}(U_L^\dagger g^f_L U_L)_{kl} m_{f_l}
\left( (1-u)m^2_{\phi_i} +\frac{u}{2}(m^2_{f_k}+m^2_{f_l}) - 
\frac{v}{2}(m^2_{f_k}-m^2_{f_l})\right)^{-1}\nonumber\\
&&-\,\delta_{kl}(U_\phi^\dagger g^\phi U_\phi)_{ij}\ln\left(
\frac{u}{2}(m^2_{\phi_i}+m^2_{\phi_j}) + (1-u) m^2_{f_k} - 
\frac{v}{2} (m^2_{\phi_i}-m^2_{\phi_j})\right) \Bigr\}.
\eea
Although it is hard to glean much from this formula by inspection,
there are three enlightening special cases where the result again
becomes simple.

(1) {No mixing.}  If $\mu$, $M_2\gg m_W$ in the chargino mass matrix
and $m_t A_t$, $m_t\mu\ll m^2_t+ m^2_{Q,U}$ in the squark mass matrix
there would be small mixing angles in both sectors.  Then the simple
formula (\ref{result}) applies, using the 1-1 elements of $g^f_L$ and
$g^f_R$, since these are the states that are defined to couple to the
left-handed $b$ quark in the absence of mixing.  In the SUSY case we
see immediately that $\delta g^b_L$ is zero in this limit, since
$g^f_L-g^f_R=0$.

(2) Mixing, but $m_{\phi_1} = m_{\phi_2}$ and $m_{f_1} = m_{f_2}$.
(Degenerate masses imply no mixing in the scalar sector, but there is
generally still mixing of the fermions because the asymmetry of the
Dirac mass matrix.)  Then eq.~(\ref{result}) again applies, but with
the replacement
\be
	g^f_L-g^f_R \to \left( U_R U_L^\dagger\, g^f_L\, U_L U_R^\dagger
	- g^f_R\right)_{11}.
\ee
With maximal mixing so that $U_R U_L = ( {0\atop 1}
\ {1\atop 0} )$, we would have $U_R U_L^\dagger\, g^f_L\, U_L U_R^\dagger
= ( {0\atop 1}\ {1\atop 0} )( {-1\atop \ze}\ {\ze\atop-\frac12} )
( {0\atop 1}\ {1\atop 0} ) = ( {-\frac12\atop \ze}\ {\ze\atop-1} )$ so
that 
\be
\left( U_R U_L^\dagger\, g^f_L\, U_L U_R^\dagger
	- g^f_R\right)_{11} = -1 +\frac12 = -\frac12,
\ee
which has the correct sign to increase $R_b$!  Thus we see that large
mixing in the chargino sector is desirable for boosting $R_b$.  The
condition can be written in terms of the mixing angles as
\bea
U_R\, U_L^\dagger &=& 
\left( \begin{array}{rr} \cos\theta_{R} & \sin\theta_{R}\\
	-\sin\theta_{R} & \cos\theta_{R} \end{array}\right)
\left( \begin{array}{rr} \cos\theta_{L} & -\sin\theta_{L}\\
	\sin\theta_{L} & \cos\theta_{L} \end{array}\right)
= \left( \begin{array}{cc} 0 & 1\\
	1& 0 \end{array}\right)\nonumber\\
	&\Longrightarrow& \tan\theta_R\, \tan\theta_L = -1\
\label{cond}
\eea

It is straightforward to work out the consequences of this restriction
for the chargino mass matrix, eq.~(\ref{chargino}).  One finds that
light charginos are required, such that $m_{\chi_1} + m_{\chi_2} \cong
2 m_W$.  Furthermore $\tan\beta$ (the ratio of the two Higgs VEV's
$\langle H_2\rangle/\langle H_1\rangle$) is restricted to lie in the
range $m_{\chi_1}/m_{\chi_2} \le \tan\beta \le m_{\chi_2}/m_{\chi_1}$.
Given the restriction on the sum of the chargino masses, their ratio
can never be very large or very small without having one of them be 
lighter than the experimental limit, and thus it is necessary to have
$\tan\beta\sim 1$.  In addition, the shift in $R_b$ is proportional to
the function ${\cal F}$ of eq.~(\ref{result}), which will suppress the
result if $\tilde t_R$ is much heavier than the charginos.  Thus
we also see that a relatively light right-handed top squark is needed.

(3) Heavy scalars.  In the limit that $m_{\phi_1},m_{\phi_2}\gg 
m_{f_1}, m_{f_2}$ the expression for $\delta g^b_L$ again simplifies:
\be
  \delta g^b_L = {y^2 n_c\over 64 \pi^2}\sin^2\! 2\theta_\phi
\left( g^\phi_{22} - g^\phi_{11} \right)
	{\cal F}'(m^2_{\phi_1}/m^2_{\phi_2})
\ee
where $\theta_\phi$ is the scalar mixing angle and ${\cal F}'$ is another 
nonnegative function.  All of the factors in this equation are thus 
positive with the possible exception of $g^\phi_{22} - g^\phi_{11}$, hence the
sign of the latter determines the sign of $\delta g^b_L$.  In the SUSY
case $g^\phi_{22} - g^\phi_{11} = \frac12$, which has the wrong sign.  Thus
one cannot have both top squarks much heavier than the charginos, in
agreement with the conclusion of the previous paragraph.

 \begin{figure} 
\epsfysize=3in\epsfbox{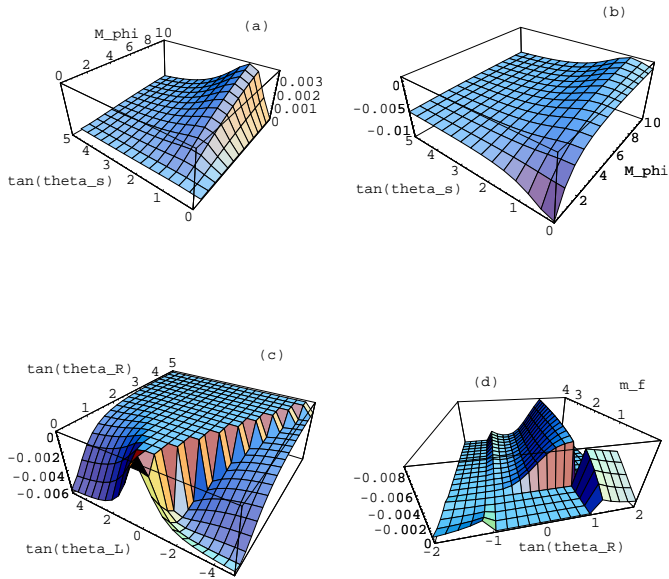} 
\caption{Dependence of $\delta g^b_L$ on various parameters in 
supersymmetric models.
\label{fig:susy}}
\end{figure}

Finally I show some quantitative predictions for $\delta g^b_L$ without
going to the any of the three limits just discussed.  
To completely account for the $R_b$ discrepancy, one wants a shift of
$\delta g^b_L = -0.0063$.  Figure 4 shows how $\delta g^b_L$
depends on various pairs of parameters in supersymmetric extensions of
the standard model.  In figure 4(a) we vary the right-handed top squark
masses and the squark mixing angle, keeping all other masses fixed and
degenerate, but with the wrong sign, $\tan\theta_L \tan\theta_R = +1$,
relative to condition (\ref{cond}).  One sees that the sign of $\delta
g^b_L$ is opposite to that needed for increasing $R_b$, in accordance
with the limiting case (2) above.  Figure 4(b) shows the same thing
except with condition (\ref{cond}) fulfilled, so that
$\delta g^b_L$ indeed has the desired sign, and also a large enough
magnitude if $\tilde t_R$ is sufficiently light\,\footnote{Since we set
$m_Z=0$, the scale of mass is set by that of the charginos.  Thus a
mass of 1 on figure 4 corresponds to the squarks and charginos all
being degenerate with each other.} or has large enough mixing.  Figure
4(c) shows that large negative values of $\delta g^b_L$ occur for
values of the chargino mixing angles in accordance with (\ref{cond}).
However we set $\delta g^b_L=0$ in the figure for any parameters giving
theoretically disfavored values of $\tan\beta<1$, which is why there are
no large values of $\delta g^b_L$ when $\tan\theta_R$ becomes large, as
eq.~(\ref{cond}) taken by itself would have predicted.  Figure 4(a)
also shows the preference for $\tan\theta_L \tan\theta_R = -1$, since
here we have set $\tan\theta_L=-1$ and allowed $\tan\theta_R$ to vary.
The largest shifts in $\delta g^b_L$ occur when $\tan\theta_R = +1$.
In addition, the mass of the mostly-Wino component of the chargino is 
allowed to vary, showing that the shift in $\delta g^b_L$ is sufficiently
large as long as the former is not much lighter than the other chargino.

\section{Conclusions}

In summary, we have found simple, analytic expressions for the shift in
the coupling of the left-handed $b$ quark to the $Z$ boson, induced by
one-loop diagrams due to new particles beyond the standard model.
These expressions make it easy to understand which models are likely to
be able to resolve the $R_b$ problem, and to identify the favored
regions of parameter space.  For example our formulas immediately show
that in the simplest supersymmetric extensions of the standard model,
one needs $\tan\beta\sim 1$, and light charginos and right-handed top
squarks, with masses near $m_W$.  This agrees with and, we hope,
illuminates the results of numerous more detailed studies.\cite{susy}
Moreover our results make it easier to see how to construct models that
can do the job the more effectively, should the need arise.  For
example, if the favored regions of SUSY parameter space mentioned above
should be ruled out, one could loosen the constraints on the chargino
masses by providing new sources of off-diagonal terms in the chargino
mass matrix that would be larger than the weak scale.  This would allow
the charginos to have large mixing angles without requiring them to be
light, and give more freedom in the choice of $\tan\beta$.

\section*{Acknowledgments} 
Our work received financial support from NSERC of Canada and FCAR du
Qu\'ebec.

\end{document}